\documentclass[twocolumn, prb, showpacs, amsmath, amssymb, superscriptaddress]{revtex4}

\usepackage{graphicx}

\begin{document}
\title{Dynamics of Fluctuations of Localized Waves}

\author{J. Wang}
\affiliation{Department of Physics, Queens College, The City University of New York, Flushing, NY 11367, USA}
\author{A. A. Chabanov}
\affiliation{Department of Physics and Astronomy, University of Texas at San Antonio, San Antonio, Texas 78249, USA}
\author{D. Y. Lu}
\affiliation{Department of Physics, Hong Kong University of Science and Technology, Clear Water Bay, Kowloon, Hong Kong}
\author{Z. Q. Zhang}
\affiliation{Department of Physics, Hong Kong University of Science and Technology, Clear Water Bay, Kowloon, Hong Kong}
\author{A. Z. Genack}
\affiliation{Department of Physics, Queens College, The City University of New York, Flushing, NY 11367, USA}
\date{\today}

\begin{abstract}
We follow the temporal evolution of mesoscopic intensity fluctuations and correlation in strongly localized samples. We find an initial burst in relative transmission fluctuations in random one-dimensional (1D) samples due to fluctuations in the arrival time of ballistic transmission. Relative fluctuations subsequently rise, then drop to a minimum at a time $t_m$, after which they increase rapidly in 1D simulations and quasi-1D (Q1D) measurements. For $t>3t_m$, results in 1D and Q1D samples converge towards predictions of a dynamic single parameter scaling model. These results reflect the changing number of modes participating appreciably in transmission as the impact of longer lived modes grows with time delay.
\end{abstract}
\maketitle
The scaling of the electronic conductance or of the transmission of classical waves characterizes the nature of transport\cite{Abrahams, SPS} and provides a window on the underlying quasimodes of the random sample\cite{Thouless, Azbel}. Such quasimodes, which we will refer to as ``modes'', correspond to resonances of an open system. An inverse variation of transmission with sample thickness indicates that transport is Ohmic or diffusive. In such samples, modes extend throughout the sample and overlap spectrally. In contrast, when transmission scales exponentially, modes are exponentially localized within the sample and isolated spectrally. The description of the transport of electronic and classical waves are strikingly similar. This can be seen in the equivalence expressed in the Landauer relation\cite {Landauer} between the dimensionless conductance $\textsl{g} \equiv G/(e^2/h)$, where $G$ is the conductance, $e$ is the electron charge, $h$ is Planck's constant, and the optical transmittance, $T$, $\textsl{g}=T \equiv \sum_{ab}T_{ab}$. Here, $T_{ab}$ is the intensity or transmission coefficient of the input and output channels $a$ and $b$, respectively. These channels can be either transverse propagation modes external to the sample or positions within distinct coherence areas on the sample surface with specified polarization. The scaling of $\textsl{g}$ depends only upon the value of $\textsl{g}$ itself and the dimensionality of the sample. Anderson localization is achieved when $\textsl{g}<1$\cite{SPS}. Localization of classical waves is of special interest because waves can be localized purely by wave interference without the complicating role of the Coulomb interaction. In addition, measurements of fluctuations of relative transmission, which are greatly enhanced in the localization transition, can be made for single channels or for sums over channels in a random ensemble of statistically equivalent samples. Indeed, $\textsl{g}$ and the variance of fluctuations of total transmission normalized by its ensemble average are inversely related\cite {Mello,Feng, Pnini,Albada, StructureCorr,Nieuwenhuizen,Kogan,Stoytchev, MaretUCF, Nature}. Thus the statistics of steady state transmission, which can be measured for classical waves, characterize propagation and localization in random media. 

Recently there has been great interest in the dynamics in the localization transition. In contrast to measurements of the scaling of transmission which can track the changing impact of weak localization on samples of different sizes, pulsed measurements may provide the changing contributions of underlying electromagnetic modes with different decay rates in samples of a particular scale. A slowdown of the rate of decay of transmission of classical waves has been observed in microwave\cite{Breakdown, PRB}, optical\cite{Maretpulse} and ultrasound\cite{Page} experiments near and beyond the localization threshold. A related reduction has been observed in the spreading of matter waves due to Anderson localization of Bose-Einstein condensates in a random potential created by 1D optical speckle patterns \cite {ColdAtom}. The ensemble average of pulsed transmission was measured for localized ultrasound just beyond the mobility edge in a slab of sintered aluminum beads\cite {Page} and for strongly localized microwave radiation in random dielectric Q1D samples\cite {PRB}. The ultrasound measurements of ensemble averaged intensity, $\langle I(t) \rangle$, were well fit by the self-consistent theory of localization (SCLT)\cite {Page, SCLT, Bart} with a renormalized diffusion coefficient in space and frequency. In strongly localized Q1D samples, however, the decay rate of microwave transmission fell below predictions for SCLT\cite{PRB}. The slowing down of transport at long times was explained in terms of a 1D dynamic single parameter scaling (DSPS) model which neglects mode overlap and averages over the distribution of decay rates and associated transmission strengths of localized modes\cite{PRB,SPS}. The relative contributions of long-lived localized modes and short-lived ``necklace states'' is central to understanding dynamics\cite{Mott,Pendry,Wiersma,Sebbah}. Necklace states are formed by the hybridization of spectrally overlapping localized states to form multiply-peaked modes in space which decay relatively rapidly through the sample boundary. In addition to a decay of average transmission, observations of a corresponding growth in correlation with time have been made in microwave measurements of in diffusive samples\cite{dynamicCorrDiff} but dynamic measurements of mesoscopic phenomena have not as yet been carried out in localized samples.

Here we report the first studies of the dynamics of mesoscopic fluctuations and correlation for localized waves. 1D simulations and microwave measurements in Q1D samples reveal a complex temporal variation of transmission statistics following an excitation pulse. A short jump is observed in relative fluctuations in 1D due to the variations in the speed of the ballistic wave in different configurations. Subsequently, relative fluctuations rise and then drop to a minimum at a time $t_m$ and then rise again. The variation of fluctuations is explained in terms of the changing effective number of modes that contribute to transmission. Fluctuations are enhanced at $t<t_m$ and $t>t_m$ by the selective contribution to transmission of short- and long-lived modes, respectively. For $t>3t_m$, results in 1D and Q1D samples converge towards predictions of the DSPS model for isolated localized modes.

Measurements are carried out on Q1D samples contained in 7.3-cm-diameter copper tubes of various lengths. The tubes are filled with alumina spheres of diameter 0.95 cm and refractive index 3.14 embedded in Styrofoam spheres of diameter 1.9 cm at an alumina volume filling fraction of 0.068. Measurements are carried out in a narrow frequency window from 10-10.24 GHz just above the first Mie resonance of the alumina spheres in which localization is fostered by near resonant scattering and a dip in the density of states\cite{PRL2001}. Spectra of the transmitted field are obtained using a vector network analyzer. Ensembles of sample realizations are created by momentarily rotating the sample tube between measurements. The average intensity localization length in the sample is $\bar\xi=30\text{ cm}$ \cite{PRB}.

The response to a Gaussian incident field pulse, $E_0(t)\sim\exp(-t^2/2\sigma^2_t)\exp(i2\pi\nu_0t)$ is obtained by taking the Fourier transform of the product of the spectra of the transmitted field and the incident pulse, $E(\nu)\sim\exp(-(\nu-\nu_0)^2/2\sigma^2)$, where $\sigma=(2\pi\sigma_t)^{-1}$. This gives the time-dependent field $E(t)$, and intensity $\left| E(t)\right|^2$. Measurements are made in a number of experimental configurations. In one, transmission of a plane wave produced by a horn antenna is detected by a 4-mm wire antenna translated on a 2-mm grid over the output surface of 61-cm-long samples in 200 sample configurations. In another experimental arrangement, waves are launched and detected using conical horns at a distance from the sample.

\begin{figure}[ht]
\includegraphics[width=3.7in]{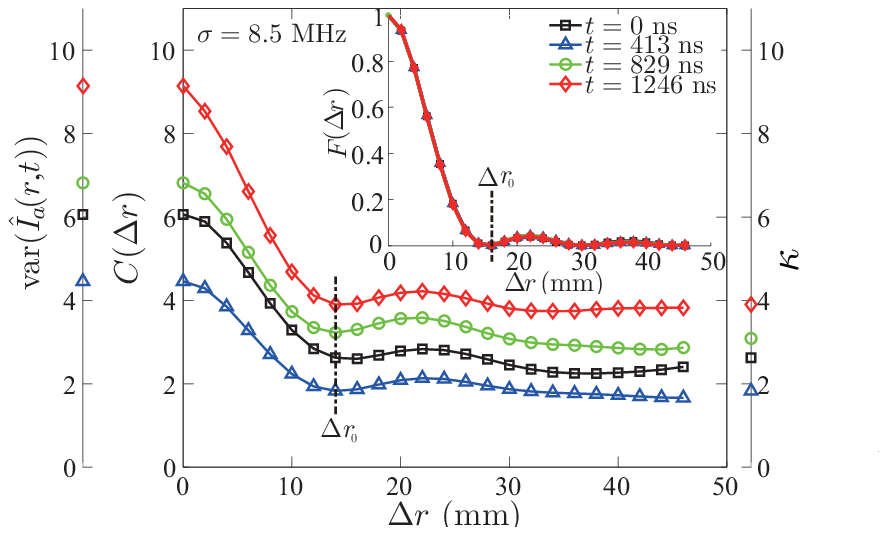}
\caption{\label{Fig1}(Color online) Spatial correlation, $C(\Delta r)$ for different delay times $t$ from the peak of an incident Gaussian pulse in a sample of $L=61\text{ cm}\sim 2\bar\xi$. $C(\Delta r)$ falls at early times and then rises, while the square of corresponding field correlation functions, $F(\Delta (r)$, shown in the inset, are independent of time. The dashed vertical lines at $\Delta r_0=12$ indicate the first zero of $F$ at which point, $C=\kappa$. $C(0)={\rm var}(\hat I_a(r))$ and $C(\Delta r_0) \equiv \kappa$ are shown on the vertical lines on the right and left sides of the figure, respectively.} 
\end{figure}
Cumulant correlation functions of relative intensity, $\hat I_a(r,t)=I_a(r,t)/\langle I_a(r,t)\rangle$, with displacement on the output surface, $\Delta r$, $C(\Delta r,t)=<\delta \hat I_a(r,t)\delta \hat I_a(r+\Delta r,t)>$, for a single incident transverse mode, $a$, are shown in Fig. \ref{Fig1} for different time delays, $t$, in Q1D samples with $L \sim 2\bar\xi$. Here $\delta \hat I_a(r)= \hat I_a(r)-1 $ is the deviation of relative intensity from its ensemble average of 1 and $\bar \xi=30\text{ cm}$ \cite{PRB} is the intensity localization length. The correlation function has the same form for localized and diffusive waves, $C(\Delta r)=F(\Delta r)+\kappa(1+F(\Delta r))$ \cite{Mello,StructureCorr}. Here $F(\Delta r, t)=|F_{E}|^2$ is the square of the normalized field correlation function and $\kappa$ is the degree of relative intensity correlation at displacements for which $F=0$, so that, for example,  $\kappa= C(\Delta r_0)$ at the first zero of $C$. In the diffusive limit, $\kappa\to 0$ and $C=F$ is correlated over the short range of a speckle spot. $F_E$ is the Fourier transform of the normalized specific intensity, which is the angular distribution of transmitted intensity, which does not change in time. As a result, $F$ is independent of $t$, as seen in the inset in Fig. \ref{Fig1}. $\kappa$ is seen to drop at early times before increasing. 
\begin{figure}[ht]
\includegraphics[width=2.8in]{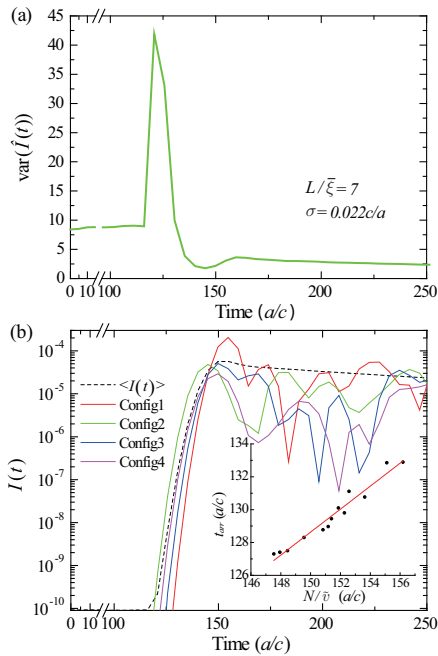}
\caption{\label{Fig2}(Color online) (a) Variance of relative intensity vs. $t$ in 1D. (b) Initial jump in ${\rm var}(\hat I(t))$ is associated with the different times of arrival $t_{arr}$ of the rising edge of $I(t)$ in different random configurations. The dashed curve is the configuration average of $I(t)$. The inset shows the arrival time $t_{arr}$ in different configurations closely matches the calculated ballistic time through the sample with $N$ elements of thickness $a$ and average phase velocity $\tilde v$.}
\end{figure}

It was not possible to study early time dynamics of correlation in Q1D samples with short pulses because of the narrow frequency band over which the statistics of propagation are uniform. However, the response of narrow pulses could be calculated in 1D simulations. Configurations of random samples are constructed using the model described in Ref. 16. Samples of $N=L/a$ layers of equal thickness $a$, are embedded in vacuum with $\epsilon =1$ and wave speed $c$, the speed of light. The dielectric constant in each layer, $\epsilon_i$, is a uniformly distributed random number between 0.3 and 1.7. The intensity localization length is $\bar\xi=-L/\langle\ln T\rangle=22a$ and the central frequency is $\nu_0=0.26c/a$\cite{PRB}.

The temporal variation of relative fluctuations in transmission in 1D following a Gaussian incident pulse, ${\rm var}(\hat I(t))$, calculated using spectra of the transmitted field just beyond the output surface for an ensemble of 50,000 configurations is shown in Fig. \ref{Fig2}(a). The sharp spike in ${\rm var}(\hat I(t))$ (Fig. \ref{Fig2}(a)) is due to differences in the transit time of the leading edge of the ballistic pulse in different random configurations (Fig. \ref{Fig2}(b)), which leads to large fluctuations in $\hat I(t)$. The time at which the leading edge of the pulse arrives at the sample output, $t_{\mathrm{arr}}$, determined by the time at which $I(t)$ reaches the value $10^{-8}$, is seen to closely track the calculated ballistic arrival time, $\sum_{i}^{N}a\sqrt{\epsilon_i}/c \equiv (N/ \tilde{v})(a/c) $. Once the leading edge of the pulse is transmitted, $I(t)$, remains high for a time equal to the inverse bandwidth of the exciting pulse and relative fluctuations drop to a minimum. After this time, the fluctuations from one configuration to another have relative maxima at different times as a result of the random phasing of contribution to transmission of modes of electromagnetic radiation at different frequencies in different sample configurations. The variance of fluctuations is then inversely proportional to the effective number of modes contributing significantly to transmission. At first, only the shortest lived modes contribute appreciably to transmission since longer-lived modes surrender their energy slowly. This leads to a peak in ${\rm var}(\hat I(t))$  at $\sim160a/c$. After this time, the contribution of longer lived modes begins to be felt since energy in the shortest lived modes has already leaked from the sample. As a result, the effective number of modes contributing to transmission increases and ${\rm var}(\hat I(t))$ falls. 

\begin{figure}[t]
\includegraphics[width=2.8in]{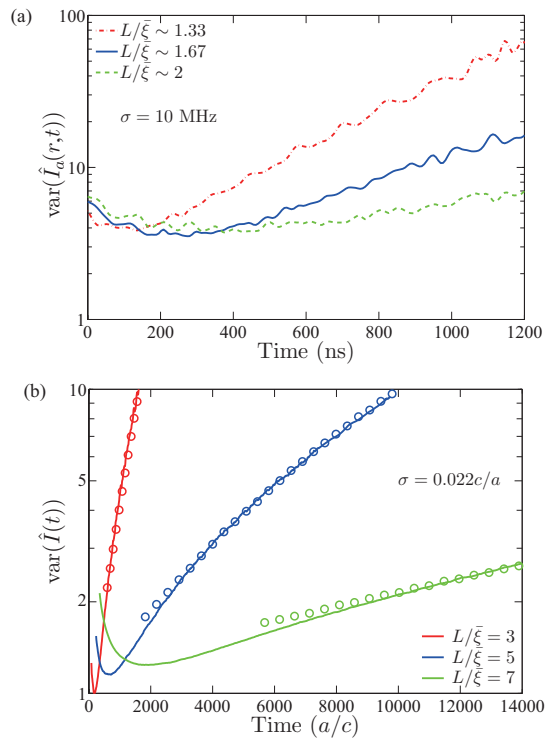}
\caption{\label{Fig3}(Color online) Dynamic variance of normalized transmission for three sample lengths and a fixed pulse width obtained from (a) measurements in Q1D for three sample lengths, $L=40\text{ cm}\sim 1.33\bar\xi$, $L=50\text{ cm}\sim 1.67\bar\xi$ and $L=61\text{ cm}\sim 2\bar\xi$, and (b) simulations in 1D (curves) and the results of the DSPS model for $t>2.5t_m$ (open circles).}
\end{figure}

\begin{figure}[t]
\includegraphics[width=2.8in]{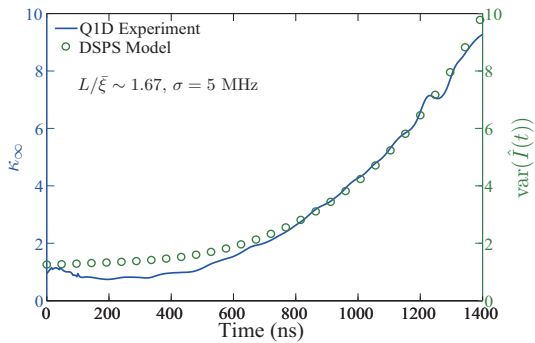}
\caption{\label{Fig4}(Color online) Comparison of measurements of the dynamics of infinite-range correlation in a Q1D sample of $L=50\text{ cm}\sim1.67\bar\xi$ and calculations of the variance of intensity in the DSPS model for a corresponding 1D sample.}
\end{figure}

The dynamics of fluctuations over a broader time scale for measurements in Q1D and simulations in 1D are shown in Figs. \ref{Fig3}-\ref{Fig4}. Measurement in quasi-1D and simulation in 1D show minima in ${\rm var}(\hat I_{a}(t))$ (Fig. \ref{Fig3}(a)) and  ${\rm var}(\hat I(t))$ (Fig. \ref{Fig3}(b)) at a time $t_m$, which we find is independent of pulse bandwidth. This indicates that $t_m$ reflects a property of modes of the medium. The minimum is associated with a transition from a condition in which short-lived modes dominate transmission to one in which long-lived modes predominate. Presumably, $t_m$ corresponds to the time within the intermediate time range of Fig. \ref{Fig3} at which the largest number of modes participate appreciably to transmission.

Since the rapid increase of $\mathrm{var}(\hat I_a(r,t)))$ after $t_m$ arises from the dominance of long-lived modes, its behavior might be modeled by the DSPS model proposed in Ref. 16. According to this model, localized modes peaked in space at depth $z$ into the sample have an amplitude $A(\gamma,z)=\exp\left(-\gamma \left|L-2z \right| \right)$ at the output surface, where the Lyapunov exponent, $\gamma=1/2\xi$, is drawn from a Gaussian distribution, $P(\gamma)=\sqrt{L/2\pi \bar \gamma}\exp \left[-(\gamma-\bar \gamma)^2/(2\bar \gamma/L)\right]$\cite{SPS} with a lower cutoff limited by the sample length, i.e., $\gamma\geq B/L$, where $B$ is the only adjustable parameter in our model. The time response to a Gaussian incident pulse at long times can be written as, 
\begin{eqnarray}
I_{\rm DSPS}(t)  =&& \left|\sum_{i=1}^ME(\nu_i)A(\gamma_i,z_i)\frac{\Gamma(\gamma_i,z_i)}{2} \right. \nonumber \\
              &&   \left. \exp\left(-\frac{\Gamma(\gamma_i,z_i)}{2}t-i2\pi\nu_it\right)\right|^2,\label{eq1}
\end{eqnarray}
where $\Gamma(\gamma_i,z_i)$ is the decay rate of a localized state located at $z_i$ with a localization length $\xi_i=1/2\gamma_i$ and $\nu_i$ is the frequency of the localized state. The explicit expression for $\Gamma(\gamma_i,z_i)$ is given in Eq. (3) of Ref. 16. In our Monte Carlo simulations of the DSPS model for $I_{\rm DSPS}(t)$ , we average over a window of size $\Delta\nu=0.207c/a$. We assume that the number of states $M$ excited inside the window follows the Poisson distribution with a mean $\bar M$  equals to $\Delta\nu\rho L$ , where the density of states per unit length at $\nu_0$ is $\rho=1.32c^{-1}$, and the frequencies $\nu_i$ ($i=1,2,\cdots,M$) are chosen randomly inside the window.  The results of the DSPS model for $t>2.5t_m$ are shown as circles in Fig. \ref{Fig3}(b). Excellent agreement between the DSPS model and 1D simulations is found for $t>3t_m$. Variances of intensity are seen in Fig. \ref{Fig3} to be larger for longer samples at early times as is found in steady state. But at later times, variances are larger in shorter samples. In this case, the density of trajectories at a given time is larger than in longer samples and the probability of trajectories crossing is greater.

Though $\langle I \rangle$ may be compared to either $\langle I_{a}(r) \rangle$, $\langle I_a\equiv \sum_rI_a(r) \rangle$ or $\langle T\equiv\sum_aT_a \rangle$ since these are the same in 1D, the strength of second order statistics in Q1D depends upon the extent of spatial averaging over the speckle pattern. The most apt comparison of ${\rm var}(\hat I)$ to second order transmission statistics in Q1D is to ${\rm var}(s=T/\langle T \rangle)$, which is equal to the infinite-range correlator, $\kappa_\infty\equiv\langle\delta \hat I_a(r)\delta \hat I_{a'}(r')\rangle$\cite{Mello,Feng,StructureCorr,Nieuwenhuizen,Kogan, MaretUCF}, where $a \neq a'$, and $r$ and $r'$ are two position on the output surface at which field correlation vanishes. Short-range correlation of the speckle pattern does not contribute to either ${\rm var}(\hat I)$ or $\kappa_\infty$. On the one hand, $\kappa_\infty$ is independent of the choice of input and output transverse mode or position in Q1D, while on the other, there is no transverse intensity variation in 1D. A DSPS calculation of ${\rm var}(\hat I)$ and measurements of $\kappa_\infty$ in Q1D for a sample with $L/\bar \xi \sim 1.67$ excited by a pulse of width $\sigma = 5\text{ MHz}$ are compared in Fig. \ref{Fig4}.
Good agreement is obtained for $t>700\text{ ns}\sim 3t_m$. We use the same values for the parameters which appear in Eq. (3) of Ref. 16, which were obtained there by fitting the decay rate at long times. The density of states per unit length in this system is $\rho=8.67\text{ ns/cm}$. These results demonstrate that for long times, measurements in Q1D approach Monte Carlo simulations of the 1D DSPS theory in corresponding samples. This convergence even in samples for which $L/\bar\xi$ is not much larger than unity reflects the similar probability distributions for $\hat I$ and $s$ for long times. This is in contrast to the log-normal distribution of $s$ in steady state predicted only for $L/\bar\xi\gg 1$ \cite{Imry,Anderson80}, and reflects the dominance of localized modes at long times.

In conclusion, the dynamics of mesoscopic fluctuations of localized waves provides a window on the evolving contributions of short- and long-lived electromagnetic modes of the random medium. For $t<t_m$ but somewhat greater than $t_{\mathrm{arr}}$, the transmitted energy is due to modes which release their energy quickly, while for $t>t_m$ a decreasing subset of long-lived modes contribute substantially to transmission leading to increasingly enhanced mesoscopic fluctuations. At $t_m$, the contributions to transmission of necklace states and long-lived localized modes are most democratically represented and the variance of fluctuations is at a minimum. These results show that complex mesoscopic transport phenomena for localized waves can be clarified by applying a modal analysis. 

This research is sponsored by the National Science Foundation (DMR0907285 and ECCS0926035) and by the Hong Kong RGC under grant number 604506.

\end{document}